\documentclass[twocolumn,showpacs,prl,superscriptaddress]{revtex4-1}
\usepackage{amsfonts}
\usepackage{amsmath}
\usepackage{amssymb}
\usepackage{graphicx}

\input{tcilatex}
\begin{document}

\title{Accurate computation of low-temperature thermodynamics for quantum
spin chains}
\date{\today }
\author{Yu-Kun Huang}
\email{ykln@mail.njtc.edu.tw}
\affiliation{Graduate school of Engineering Science and Technology,\\
Nan Jeon Institute of Technology, Tainan 73746, Taiwan}
\author{Pochung Chen}
\email{pcchen@phys.nthu.edu.tw}
\affiliation{Department of Physics and Frontier Research Center on Fundamental and
Applied Sciences of Matters, \\
National Tsing Hua University, Hsinchu 30013, Taiwan}
\author{Ying-Jer Kao}
\email{yjkao@phys.ntu.edu.tw}
\affiliation{Department of Physics and Center for Advanced Study of Theoretical Science,\\
National Taiwan University, No. 1, Sec. 4, Roosevelt Rd., Taipei 10607,
Taiwan}

\begin{abstract}
We apply the biorthonormal transfer-matrix renormalization group (BTMRG)
[Phys. Rev. E \textbf{83}, 036702 (2011)] to study low-temperature
properties of quantum spin chains. Simulation on isotropic Heisenberg spin-$%
1/2$ chain demonstrates that the BTMRG outperforms the conventional
transfer-matrix renormalization group (TMRG) by successfully accessing far
lower temperature unreachable by conventional TMRG, while retaining the same
level of accuracy. The power of the method is further illustrated by the
calculation of the low-temperature specific heat for a frustrated spin
chain. 

\end{abstract}

\maketitle

Quasi-one-dimensional (Q1D) quantum spin systems have been the focus of
intensive research for the past decades. Quantum fluctuation plays an
important role in these systems, and powerful non-perturbative analytical
methods are 
available (See, for example, Ref.~\cite{Giamarchi:2004cr}). Recently, there
is a resurgence of interests in Q1D magnetic materials with spin spiral
states such as Rb$_{2}$Cu$_{2}$Mo$_{3}$O$_{12}$ \cite{Hase:2004dq}, LiCuVO$%
_{4}$ \cite{Enderle:2005zr}, Li$_{2}$ZrCuO$_{4}$\cite{Drechsler:2007ve}, and
LiCu$_{2}$O$_{2}$\cite{Park:2007yq}, due to their close association with
multiferroicity \cite{Cheong:2007fk}. Typically these systems are frustrated
and Jordan-Wigner transformation approaches can only be applied in limited
cases. Consequently numerical methods become the major tools for
understanding these systems. It is well known that the density matrix
renormalization group (DMRG) \cite%
{SWhite1992,*SWhite1993,*Schollwock:2005qf,*Hallberg:2006bh} is the most
powerful numerical method to study ground state properties of Q1D strongly
correlated lattice models with extremely high precision. DMRG is further
developed into the transfer matrix renormalization group (TMRG) to study
thermodynamics at finite temperature by mapping a 1D quantum system onto a
2D classical counterpart and representing the partition function as the
trace of powers of quantum transfer matrices (QTMs) \cite%
{Bursill:1996vn,*XWang1997}. TMRG has been applied to study a variety of
quantum spin chain systems \cite%
{TXiang1998,*Naef:1999nx,*Sirker:2002kl,*Sirker:2004oq,Lu:2006fk}. In
particular, studies on frustrated Q1D spin chains have predicted several
exotic quantum phases and very rich phase diagrams. However, the behavior at
very low temperature remains difficult to study \cite{Lu:2006fk,*Sota:2010fk}%
. This is because the conventional TMRG suffers from numerical instabilities
at low temperatures due to the difficulties in accurately determining the
eigenvalues and eigenvectors of the non-Hermitian reduced density matrix~%
\cite{XWang1997}. Improved TMRG schemes which can access low temperature
regime are hence called for.




In this \textit{Letter}, we apply the biorthonormal TMRG (BTMRG) method \cite%
{YHuang2011a,*YHuang2011b} to accurately determine thermodynamic quantities
of 1D spin chains at low temperatures far below TMRG can reach, while
retaining the same level of accuracy. BTMRG is built upon a series of dual
biorthonormal bases for the left and right dominant eigenvectors of the
non-Hermitian QTM \cite{YHuang2011a,*YHuang2011b}. Here, the dual
biorthonormal bases indicate any two sets of vectors $\{ \left \vert {\alpha 
}\right \rangle \}$ and 
$\left \{ \left \vert {\beta }\right \rangle \right \} $ 
satisfying $\left \langle {\alpha }\right \vert {\beta }\rangle =\delta
_{\alpha \beta }$. From the results of the 1D spin-1/2 Heisenberg and
frustrated $J_{1}$-$J_{2}$ models, we believe that BTMRG can reach
temperatures that is far below TMRG can reach without losing accuracy or
suffering from numerical instability. This opens up new possibilities to
study interesting physics such as finite-temperature behaviors near a
quantum critical point, which are inaccessible previously by other numerical
methods.

\begin{figure}[tb]
\includegraphics[width=8cm,clip]{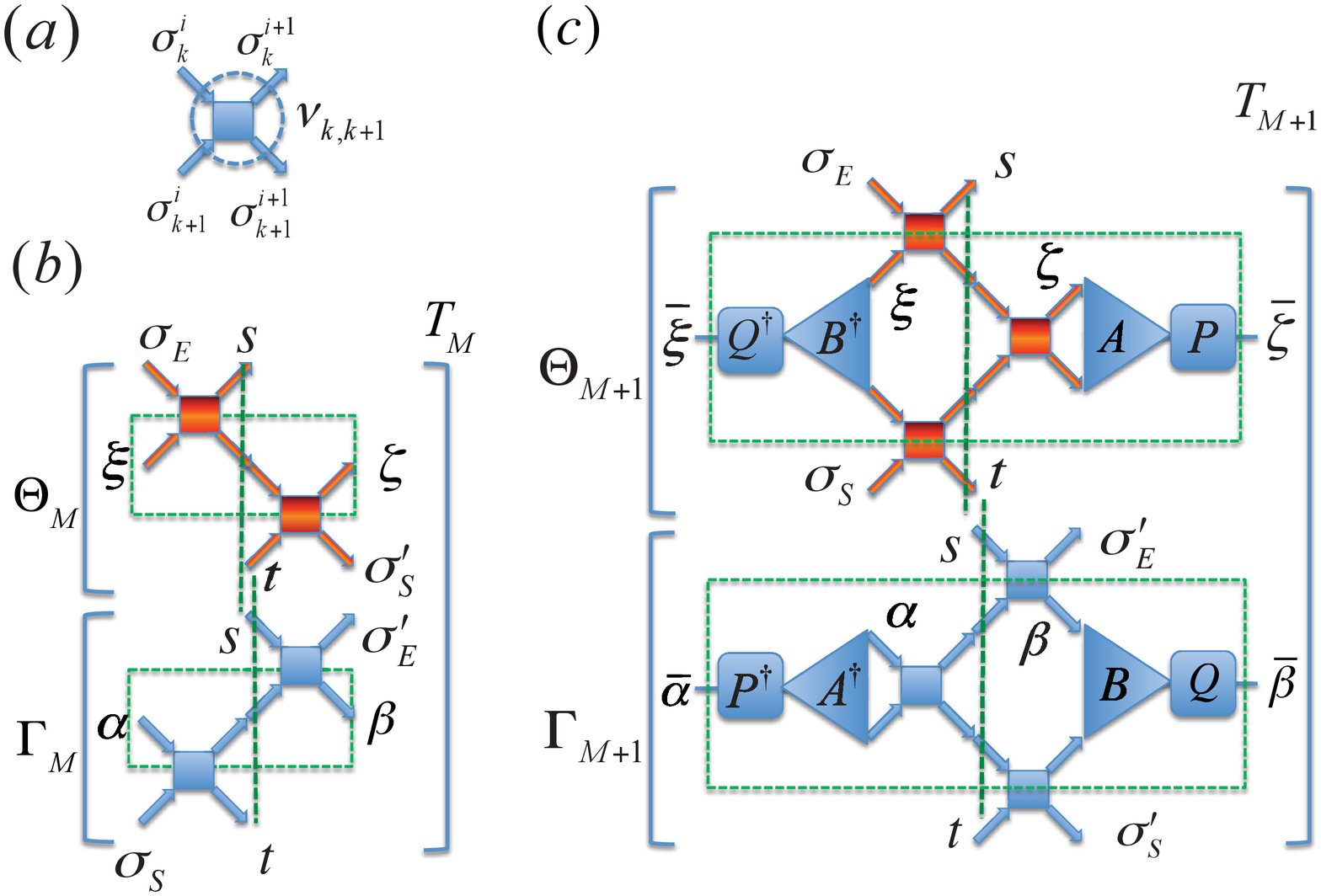} 

\vspace{-1cm} 
\caption{(Color online) (a) A single vertex in the QTM. (b) Configuration of
the augmented system (environment) block $\Gamma _{M}$ ($\Theta _{M}$),
where $\left \{ \left \vert {\protect \alpha }\right \rangle \right \} $ 
and $\left \{ \left \vert {\protect \beta }\right \rangle \right
\} $ ($\left \{
\left \vert {\protect \xi }\right \rangle \right \} $ and $\left \{
\left
\vert {\protect \zeta }\right \rangle \right \} )$ represent the
reduced dual biorthonormal bases of the current system (environment) block.
The superblock $T_{M}$ is formed by contracting indices $s$, and $t$. (c)
The enlarged augmented system (environment) block $\Gamma _{M+1}$ ($\Theta
_{M+1})$ is obtained by adding one vertex and mapping it onto the new dual
biorthonormal bases associated with $\left \{ \left \vert \overline{\protect \alpha }\right \rangle \right \}$ 
and $\left \{ \left \vert \overline{\protect \beta }\right \rangle \right \} $, 
obtained through renormalization of $T_{M}$. } \label{fig1}
\end{figure}

Let us consider the Hamiltonian ${H}$\ of a 1D quantum system of \textit{N}
(even) sites, 
\begin{eqnarray}
{H} &=&\underset{i=1}{\overset{N}{\sum }}{h}_{i,i+1}={H}_{1}+{H}_{2},  \notag
\\
{H}_{1} &=&\sum_{i=1}^{N/2}{h}_{2i-1,2i},\quad {H}_{2}=\sum_{i=1}^{N/2}{h}%
_{2i,2i+1}.
\end{eqnarray}%
Using Trotter-Suzuki decomposition \cite%
{HTrotter1958,*MSuzuki1976,*MSuzuki1985} and inserting $2M$ complete sets of
states $\{|\sigma _{k}^{i}\rangle \}$ with site index $i$ and Trotter index $%
k$, the partition function can be written as 
\begin{eqnarray}
Z &\simeq &\mathrm{Tr}\, \left \{ \left( e^{-\varepsilon {H}%
_{1}}e^{-\varepsilon {H}_{2}}\right) ^{M}\right \}  \notag \\
&=&\sum_{\left \{ \sigma _{k}^{i}\right \}
}\prod_{k=1}^{M}\prod_{i=1}^{N/2}v_{2k-1,2k}^{2i-1,2i}v_{2k,2k+1}^{2i,2i+1}=%
\text{Tr}\left( T_{M}^{N/2}\right) ,
\end{eqnarray}%
where $\varepsilon =1/MT$ and the periodic boundary conditions along both
spatial and Trotter directions are assumed. This maps a 1D quantum system
onto a 2D classical transfer-matrix tensor network. We define the QTM with
length $2M$ as $T_{M}\equiv \left( v_{1,2}v_{3,4}\cdots v_{2M-1,2M}\right)
\left( v_{2,3}v_{4,5}\cdots v_{2M,1}\right) $, where $v_{k,k+1}^{i,i+1}%
\equiv \langle \sigma _{k}^{i}\sigma _{k}^{i+1}|e^{-\epsilon
h_{i,i+1}}|\sigma _{k+1}^{i}\sigma _{k+1}^{i+1}\rangle $ (Fig.~\ref{fig1}%
(a)), and the site index can be suppressed due to the translational
invariance. Note that $T_{M}$ is \textit{real-valued} but \textit{%
non-Hermitian}. The maximum eigenvalue and the corresponding eigenvectors of 
$T_{M}$ determine all the physical properties in the thermodynamic limit. In
practice the imaginary-time step $\varepsilon $\ is usually kept fixed. For
low temperatures ($M$ large) the size of $T_{M}$ is beyond the reach of
exact diagonalization and DMRG algorithm is applied to approximately
determine the maximal eigenvalue and eigenvectors of $T_{M}$ \cite%
{Bursill:1996vn,*XWang1997}.


The heart of the BTMRG is the bi-orthonormalization procedure for any two
sets of arbitrary vectors $\left \{ \left \vert a\right \rangle \right \} $
and $\left \{ \left \vert b\right \rangle \right \} $, where $a,b=1,\cdots
,m $. Let $A\equiv \lbrack \left \vert a\right \rangle ]$\ and $B\equiv
\lbrack \left \vert b\right \rangle ]$\ be matrices with the vectors $%
\left
\vert a\right \rangle $\ and $\left \vert b\right \rangle $ as their
columns. Performing a singular value decomposition (SVD) upon $A^{\dag }B$
to obtain $A^{\dag }B=U\Lambda V^{\dag }$, we can readily obtain a dual set
of biorthonormal bases\ $\widetilde{A}\equiv \lbrack \left \vert \alpha
\right
\rangle ]=AP$\ and\ $\widetilde{B}\equiv \lbrack \left \vert \beta
\right
\rangle ]=BQ$, where\ $P=U\Lambda ^{-1/2}$\ and\ $Q=V\Lambda ^{-1/2}$%
\ represent non-unitary basis transformations. In the spirit of DMRG, as
shown in Fig.~\ref{fig1}(b), $T_{M}$ is partitioned into a system block and
an environment block (enclosed in dashed frames) together with two
additional time slices (labeled by $\sigma _{S}^{\phantom{\prime}}$, $\sigma
_{S}^{\prime }$ and $\sigma _{E}^{\phantom{\prime}},\sigma _{E}^{\prime }$).
At any step, we keep a dual set of biorthonormal bases for current system
(environment) block labeled by $\left \{ \left \vert \alpha \right \rangle
\right \} $\ and $\left \{ \left \vert \beta \right \rangle \right \} $ ($%
\left \{ \left \vert \xi \right \rangle \right \} $ and $\left \{
\left
\vert \zeta \right \rangle \right \} $) for the left and right
dominant eigenvectors.\ Matrix elements of the augmented system block ${%
\Gamma }_{M}$ are obtained by projecting ${\Gamma }_{M}$ onto the dual
biorthonormal bases associated with $\left \{ \left \vert \alpha
\right
\rangle \right \} $\ and $\left \{ \left \vert \beta \right \rangle
\right
\} $ as ${\Gamma }_{M}(s{\alpha }\sigma _{S}^{\phantom{\prime}%
},\sigma _{E}^{\prime }{\beta }t)=\left
\langle s{\alpha }\sigma
_{S}\right
\vert \Gamma _{M}\left \vert \sigma _{E}^{\prime }{\beta }%
t\right \rangle $, where $|\sigma _{E}^{\prime }\beta t\rangle \equiv
|\sigma _{E}^{\prime }\rangle \otimes |\beta \rangle \otimes |t\rangle $,
and similarly for the augmented environment block $\Theta _{M}$ onto the
bases associated with $\left \{ \left \vert {\xi }\right \rangle \right \} $%
\ and $\left \{ \left
\vert {\zeta }\right \rangle \right \} $. In the case
where the Hamiltonian has the reflection-symmetry ${h}_{i,i+1}={h}_{i+1,i}$, 
$\Theta _{M}$ is simply the reflection of ${\Gamma }_{M}$, ${\Theta }%
_{M}(\sigma _{E}^{\phantom{\prime}}{\xi }t,s{\zeta }\sigma _{S}^{\prime })={%
\Gamma }_{M}(s{\zeta }\sigma _{S}^{\prime },\sigma _{E}{\xi }t)$, such that
only the augmented system block ${\Gamma }_{M}$ needs to be calculated and
stored. The superblock $T_{M}$ is built by connecting ${\Gamma }_{M}$ and ${%
\Theta }_{M}$ through contracting the indices $s$ and $t$:%
\begin{eqnarray}
\lefteqn{T_{M}(\xi \sigma _{E}\alpha \sigma _{S},\zeta \sigma _{E}^{\prime
}\beta \sigma _{S}^{\prime })}  \notag \\
&=&\underset{s,t}{\sum }{\Gamma }_{M}(s{\alpha }\sigma _{S},\sigma
_{E}^{\prime }{\beta }t){\Theta }_{M}(\sigma _{E}{\xi }t,s{\zeta }\sigma
_{S}^{\prime }).
\end{eqnarray}


\begin{figure}[tb]
\includegraphics[width=8cm,clip]{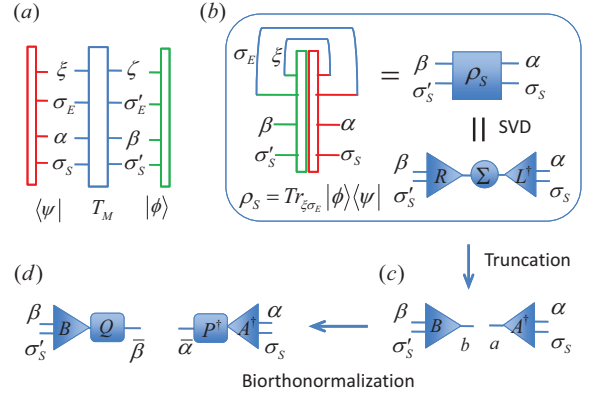} \vspace{-0.8cm} 
\caption{(Color online) Main steps of RG iteration in BTMRG. 
(a) Computation of the maximal eigenvalue and left and right eigenvectors 
$|\protect \psi \rangle $, $|\protect \phi \rangle $ of $T_{M}$. 
(b) The reduced density matrix $\protect \rho_{s}$ is formed, 
using $|\protect \psi \rangle $ and $|\protect \phi \rangle $ 
and tracing out the environment. Performing SVD on $\protect \rho_{s}$ and 
(c) keeping two reduced sets of basis vectors $\{|a\rangle \}$ and $\{|b\rangle \}$
from singular vectors of $\protect \rho _{s}$ in $L$ and $R$.
(d) The bases are biorthonormalized to form a new reduced dual biorthonormal bases 
$|\{ \overline{\protect \alpha }\rangle \}$ and $\{|\overline{\protect \beta }\rangle \}$ 
for the new system block as in Fig. 1(c). } \label{fig2}
\end{figure}


Figure~\ref{fig2} sketches the main steps of the renormalization group (RG)
iteration in BTMRG. We first calculate the maximum eigenvalue and the right
eigenvectors $\left \vert \phi \right \rangle $ of ${T}_{M}$ (Fig.~\ref{fig2}%
(a)). The left eigenvector $\left \vert \psi \right \rangle $\ can be
directly read out from $\left \vert \phi \right \rangle $: $\psi _{{\xi }%
\sigma _{E}{\alpha }\sigma _{S}}=\phi _{{\beta }\sigma _{E}{\zeta }\sigma
_{S}}$ \cite{XWang1997}. No explicit construction of ${T}_{M}$ is required
since only matrix-vector multiplication is involved in determining the
maximum eigenvalue and eigenvectors, leading to a dramatic reduction of
memory usage and computation time. The reduced biorthonormal bases $\{|%
\overline{\alpha }\rangle \}$ and $\{|\overline{\beta }\rangle \}$ for the
new system block are determined from a non-Hermitian reduced density matrix $%
{\rho }_{S}=Tr_{\sigma _{E}\xi }\left \vert \phi \right \rangle
\left
\langle \psi \right \vert$ (Fig.~\ref{fig2}(b)). By carrying out SVD
upon $\rho _{S}=R\Sigma L^{\dag }$, we keep two reduced sets of vectors $%
\left \{ \left \vert a\right \rangle \right \} $\ and $\left \{ \left \vert
b\right
\rangle \right \} $\ from singular vectors in $L$\ and $R$\
corresponding to the $m$ largest singular values in $\Sigma $ (Fig.~\ref%
{fig2}(c)). Finally, we perform the prescribed biorthonormalization
procedure to obtain the dual bases $\{|\overline{\alpha }\rangle \}$\ and $%
\{|\overline{\beta }\rangle \}$. This process is equivalent to applying
non-unitary basis transformations $P $\ and $Q$\ upon $\left \{ \left \vert
a\right \rangle \right \} $\ and $\left \{ \left \vert b\right \rangle
\right \} $ (Fig.~\ref{fig2}(d)). This completes a cycle of the RG steps in
BTMRG. The augmented system block is enlarged by adding one vertex to $%
\Gamma _{M}$, and the enlarged Hilbert space is truncated by mapping back to
the new reduced biorthonormal bases associated with $\{|\overline{\alpha }%
\rangle \}$ and $\{|\overline{\beta }\} $ (Fig.~\ref{fig1}(c)). The next
cycle is repeated with the new system block until the desired $M$ is reached.

Instead of performing SVD, conventional TMRG diagonalizes ${\rho }_{S}$\ to
obtain two reduced sets of left and right eigenvectors, which automatically
satisfy the biorthonormal condition. The BTMRG makes clear this insight and
generalizes the conventional TMRG to a broad class of biorthonormal bases.
We note that the choices of the biorthonormal bases are not unique \cite%
{YHuang2011a,*YHuang2011b} and the the steps proposed above are crucial in
order to exploit symmetries of the Hamiltonian. 
The advantage of BTMRG over the conventional TMRG is now clear. In TMRG, the
complete diagonalization of the non-Hermitian reduced density matrix will
inevitably encounter the problem of complex eigenvalues and eigenvectors,
and introduces numerical instability that prevents one from reaching very
low temperatures. In BTMRG, on the other hand, only SVD is involved and we
obtain real singular values and singular vectors. Note that the small
singular values obtained during the biorthonormaliztion procedure can also
induce instability to the BTMRG algorithm. But this can be remedied by
replacing the dual corresponding basis vectors associated with the small
singular value \cite{YHuang2011b}.

We demonstrate the power of BTMRG using isotropic Heisenberg spin-$1/2$
chain whose Hamiltonian reads: 
\begin{equation}
{H}=\sum_{i=1}^{N}J\left( {S}_{i}^{x}{S}_{i+1}^{x}+{S}_{i}^{y}{S}_{i+1}^{y}+{%
S}_{i}^{z}{S}_{i+1}^{z}\right) ,  \label{eq:Hamil}
\end{equation}%
%
%
%
%
%
%
The local Hamiltonian ${h}_{i,i+1}$ in Eq.~(\ref{eq:Hamil}) conserves total
spin thus $q_{i}=\sum_{k}(-1)^{i+k}\sigma _{k}^{i}$ can be regarded as a
good quantum number of the QTM \cite{XWang1997}. Consequently both $T_{M}$
and the reduced density matrix $\rho _{S}$ are block-diagonal and the SVD
and the biorthonormalization can both be carried out independently for each
subblock for different $q_{i}$. In addition, the maximum eigenvalue occurs
in the subblock labeled by $q_{i}=0$ \cite{Nomura:1991uq}, which results in
further simplification. The Helmholtz free energy in the thermodynamic limit
can be calculated from the maximum eigenvalue $\Lambda _{0}$ of ${T}_{M}$ as 
\begin{equation}
f=-T\underset{N\rightarrow \infty }{\lim }\frac{\log (Z)}{N}=-\frac{T}{2}%
\log \Lambda _{0}.
\end{equation}%
Other thermodynamics quantities can either be calculated from the numerical
derivatives of the free energy or directly using the dominant eigenvectors
and eigenvalues of the QTM. Systematic errors come from two sources: the
finiteness of the imaginary-time step $\varepsilon $, and truncation of the
reduced basis set. In principle, the first source of error has a $%
O(\varepsilon ^{2})$ correction in the partition function. In practice we
find that $\varepsilon =0.05$ is adequate for the models studied in this
work. On the other hand, the truncation error can be estimated by the
discarded weight $w_{d}=1-\sum_{i=1}^{m}\lambda _{i}$, where $\lambda _{i}$%
's are the singular values in the SVD of the reduced density matrix $\rho
_{s}$ and $m$ is the number of basis states kept by BTMRG. In this work, we
fixed the discarded weight to be $w_{d}=10^{-20}$ and set up a maximum
number of basis states $m$ which leads to a fast computation in the first
several hundred iterations of the RG steps.


\begin{figure}[tb]
\includegraphics[width=8cm,clip]{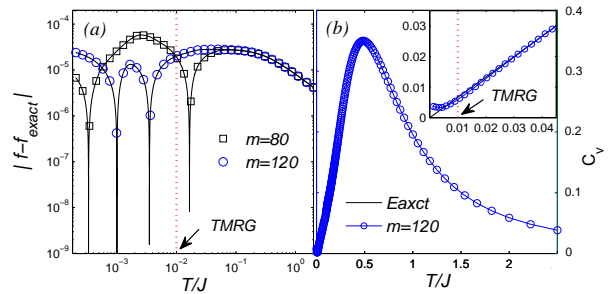} \vspace{-0.5cm} 
\caption{(Color online) 
(a) Deviations of the BTMRG free energy from the exact results for 
the isotropic Heisenberg model down to temperature $T/J=0.0002$ for both $m=80$ and $120$. 
(b) BTMRG and exact results for the specific heat $C_v$.
Inset: Low temperature regime where the exact results is $C_v=(2/3)T$.
The TMRG arrows indicate the typical temperature reachable by conventional TMRG.
} \label{fig_XY}
\end{figure}


In Fig.~\ref{fig_XY}(a) we plot the deviation of the BTMRG results with $%
m=80 $\ and $120$\ from 
the solution based on the Bethe ansatz \cite{Klumper:1993ly,Klumper:2004zr}. 
We find that the BTMRG is very stable and can reach temperature down to $%
T/J=0.0002$ for both $m$'s. For a similar $m$, conventional TMRG can only
reach down to $T/J=0.01$ due to the numerical instability. Furthermore, we
note that $T/J=0.0002$\ is not yet the limit of our algorithm, and even
lower temperature can be reached by continuing the RG iteration before
eventually the numerical instability sets in. More importantly, the accuracy
is competitive with the conventional TMRG down to temperature $T/J=0.01$\ 
\cite{Sirker:2002ys}, and the absolute error remains within the order of $%
10^{-5}$\ down to $T/J=0.0002$. At low temperatures the truncation error
dominates and higher accuracy can be reached by keeping more states, which
is clearly shown in Fig.~\ref{fig_XY}. At high temperatures, the Trotter
error due to the finite imaginary time step dominates. Due to the small
length of the QTM, the number of states kept by BTMRG to keep the discarded
weight $w_{d}=10^{-20}$\ may not reach the cutoff $m$. Consequently, the
accuracy is less sensitive to $m$. Since the calculation is not variational,
the numerical results might be larger or smaller than exact values, and an
zero error crossing will give rise to the cusps in the error curve. 
In Fig.~\ref{fig_XY}(b) we show the results of the specific heat $C_{v}$,
and an excellent agreement with the exact result is obtained. In the inset,
we zoom into the low temperature regime where the exact result is $%
C_{v}=(2/3)T$\textbf{\ }\cite{Bursill:1996vn,*XWang1997}. The BTMRG results
start to deviate from the exact value at a much lower temperature compared
to conventional TMRG. The deviation is due to accumulation of truncation
error and not the numerical instability.

\begin{figure}[tb]
\includegraphics[width=8cm,clip]{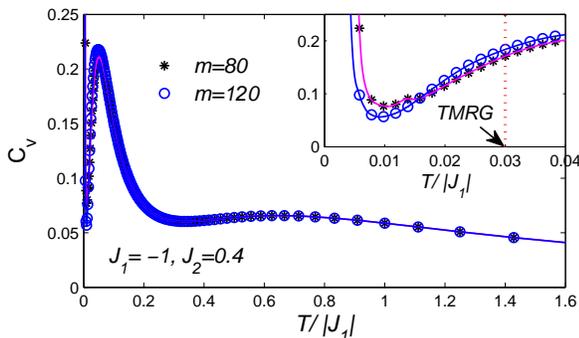} \vspace{-0.5cm} 
\caption{(Color online) Specific heat $C_v$ as a function of the temperature for
the $J_{1}$\textbf{-}$J_{2}$ model down to temperature $T/|J_{1}|=0.003$.
Inset: The inflection of the $C_v$ near below $T/|J_{1}|=0.01$ is a result of 
accumulation of truncation error.
} \label{fig4}
\end{figure}

To examine the scalability of BTMRG, we also study 
the 1D frustrated $J_{1}$-$J_{2}$ model with the Hamitonian: 
\begin{equation}
H=\sum_{i=1}^{N}J_{1}\vec{S}_{i}\cdot \vec{S}_{i+1}+J_{2}\vec{S}_{i}\cdot 
\vec{S}_{i+2}.
\end{equation}%
We consider the case of ferromagnetic nearest-neighbor (NN) interaction $%
(J_{1}=-1)$ and antiferromagnetic next-nearest-neighbor (NNN) interaction $%
(J_{2}=0.4)$. We block two adjacent spins as a superspin to obtain a new
local Hamiltonian with only NN interaction. Moreover, the QTM of this model
possesses neither reflection symmetry nor conserved quantum numbers, which
makes the computation much more numerically demanding. 
In Fig.~\ref{fig4} we calculate the specific heat $C_{v}$ down to
temperature $T/|J_{1}|=0.003$ for $m=80$ and $120$. We find a sharp
low-temperature anomaly and a broad shoulder at higher temperatures which
are in good agreement with the results of conventional TMRG \cite{Lu:2006fk}%
. In the inset of Fig.~\ref{fig4} the low-temperature behavior is magnified.
We observe an inflection of the $C_{v}$ curve at very low-temperature. For
larger $m$ the $C_{v}$ starts to inflect at lower temperature, indicating
the inflection is due to the accumulation of truncation error and not the
numerical stability. By increasing the value of $m$, the BTMRG can still be
accurately pushed to far lower temperature unreachable by TMRG without
suffering from numerical instability.


Our BTMRG computations were performed by using Matlab on a laptop with an
Intel Core i5@ 2.3GHz CPU and 4G RAM. It takes about five days to generate a
supperblock with size $2M=2 \cdot 10^5$ for $m=120$. Since BTMRG does not
involve complicated eigenvalue decomposition of the non-Hermitian $\rho_{s}$
as in conventional TMRG, no special numerical tricks are necessary, and the
computational complexity is significantly reduced.


In summary, we apply BTMRG method to accurately determine the thermodynamic
properties of 1D quantum spin chains at low temperatures. Due to the
improved numerical stability of the BTMRG, the simulations can access
temperatures several orders of magnitude lower than the conventional TMRG,
while retaining the same level of accuracy. This work suggests that the
BTMRG is a promising numerical method to investigate the low temperature
thermodynamics of Q1D quantum systems. For example, it may become possible
to study the behavior near a quantum critical point, the transition from
high-temperature regime into the quantum critical regime, and make direct
comparison with experiments. In addition it may also be possible to extend
the long-time limit in the studies of real-time dynamics \cite{Sirker:2005tg}%
.

\begin{acknowledgments}
We acknowledge the support by NSC in Taiwan through Grants No.
100-2115-M-232-001 (Y.K. Huang.), 98-2112-M-007-010 (P. Chen),
100-2112-M-002-013-MY3 (Y.J. Kao), and by NTU Grant numbers 10R80909-4 (Y.J.
Kao). Travel support from NCTS in Taiwan is also acknowledged. Y.K. Huang
would also like to give special thanks to Prof. Tao Xiang for useful
discussions.
\end{acknowledgments}

\bibliographystyle{apsrev4-1}
\bibliography{BTMRG}

\providecommand{\noopsort}[1]{}\providecommand{\singleletter}[1]{#1}%
\begin{thebibliography}{28}%
\makeatletter
\providecommand \@ifxundefined [1]{%
 \@ifx{#1\undefined}
}%
\providecommand \@ifnum [1]{%
 \ifnum #1\expandafter \@firstoftwo
 \else \expandafter \@secondoftwo
 \fi
}%
\providecommand \@ifx [1]{%
 \ifx #1\expandafter \@firstoftwo
 \else \expandafter \@secondoftwo
 \fi
}%
\providecommand \natexlab [1]{#1}%
\providecommand \enquote  [1]{``#1''}%
\providecommand \bibnamefont  [1]{#1}%
\providecommand \bibfnamefont [1]{#1}%
\providecommand \citenamefont [1]{#1}%
\providecommand \href@noop [0]{\@secondoftwo}%
\providecommand \href [0]{\begingroup \@sanitize@url \@href}%
\providecommand \@href[1]{\@@startlink{#1}\@@href}%
\providecommand \@@href[1]{\endgroup#1\@@endlink}%
\providecommand \@sanitize@url [0]{\catcode `\\12\catcode `\$12\catcode
  `\&12\catcode `\#12\catcode `\^12\catcode `\_12\catcode `\%12\relax}%
\providecommand \@@startlink[1]{}%
\providecommand \@@endlink[0]{}%
\providecommand \url  [0]{\begingroup\@sanitize@url \@url }%
\providecommand \@url [1]{\endgroup\@href {#1}{\urlprefix }}%
\providecommand \urlprefix  [0]{URL }%
\providecommand \Eprint [0]{\href }%
\providecommand \doibase [0]{http://dx.doi.org/}%
\providecommand \selectlanguage [0]{\@gobble}%
\providecommand \bibinfo  [0]{\@secondoftwo}%
\providecommand \bibfield  [0]{\@secondoftwo}%
\providecommand \translation [1]{[#1]}%
\providecommand \BibitemOpen [0]{}%
\providecommand \bibitemStop [0]{}%
\providecommand \bibitemNoStop [0]{.\EOS\space}%
\providecommand \EOS [0]{\spacefactor3000\relax}%
\providecommand \BibitemShut  [1]{\csname bibitem#1\endcsname}%
\let\auto@bib@innerbib\@empty
\bibitem [{\citenamefont {Giamarchi}(2004)}]{Giamarchi:2004cr}%
  \BibitemOpen
  \bibfield  {author} {\bibinfo {author} {\bibfnamefont {T.}~\bibnamefont
  {Giamarchi}},\ }\href
  {http://www.amazon.com/exec/obidos/redirect?tag=citeulike07-20\&path=ASIN/0198525001}
  {\emph {\bibinfo {title} {{Quantum Physics in One Dimension}}}}\ (\bibinfo
  {publisher} {Oxford University Press, USA},\ \bibinfo {year}
  {2004})\BibitemShut {NoStop}%
\bibitem [{\citenamefont {Hase}\ \emph {et~al.}(2004)\citenamefont {Hase},
  \citenamefont {Kuroe}, \citenamefont {Ozawa}, \citenamefont {Suzuki},
  \citenamefont {Kitazawa}, \citenamefont {Kido},\ and\ \citenamefont
  {Sekine}}]{Hase:2004dq}%
  \BibitemOpen
  \bibfield  {author} {\bibinfo {author} {\bibfnamefont {M.}~\bibnamefont
  {Hase}}, \bibinfo {author} {\bibfnamefont {H.}~\bibnamefont {Kuroe}},
  \bibinfo {author} {\bibfnamefont {K.}~\bibnamefont {Ozawa}}, \bibinfo
  {author} {\bibfnamefont {O.}~\bibnamefont {Suzuki}}, \bibinfo {author}
  {\bibfnamefont {H.}~\bibnamefont {Kitazawa}}, \bibinfo {author}
  {\bibfnamefont {G.}~\bibnamefont {Kido}}, \ and\ \bibinfo {author}
  {\bibfnamefont {T.}~\bibnamefont {Sekine}},\ }\href {\doibase
  10.1103/PhysRevB.70.104426} {\bibfield  {journal} {\bibinfo  {journal} {Phys.
  Rev. B}\ }\textbf {\bibinfo {volume} {70}},\ \bibinfo {pages} {104426}
  (\bibinfo {year} {2004})}\BibitemShut {NoStop}%
\bibitem [{\citenamefont {Enderle}\ \emph {et~al.}(2005)\citenamefont
  {Enderle}, \citenamefont {Mukherjee}, \citenamefont {F{\aa}k}, \citenamefont
  {Kremer}, \citenamefont {Broto}, \citenamefont {Rosner}, \citenamefont
  {Drechsler}, \citenamefont {Richter}, \citenamefont {Malek}, \citenamefont
  {Prokofiev}, \citenamefont {Assmus}, \citenamefont {Pujol}, \citenamefont
  {Raggazzoni}, \citenamefont {Rakoto}, \citenamefont {Rheinst{\"a}dter},\ and\
  \citenamefont {R{\o}nnow}}]{Enderle:2005zr}%
  \BibitemOpen
  \bibfield  {author} {\bibinfo {author} {\bibfnamefont {M.}~\bibnamefont
  {Enderle}}, \bibinfo {author} {\bibfnamefont {C.}~\bibnamefont {Mukherjee}},
  \bibinfo {author} {\bibfnamefont {B.}~\bibnamefont {F{\aa}k}}, \bibinfo
  {author} {\bibfnamefont {R.~K.}\ \bibnamefont {Kremer}}, \bibinfo {author}
  {\bibfnamefont {J.-M.}\ \bibnamefont {Broto}}, \bibinfo {author}
  {\bibfnamefont {H.}~\bibnamefont {Rosner}}, \bibinfo {author} {\bibfnamefont
  {S.-L.}\ \bibnamefont {Drechsler}}, \bibinfo {author} {\bibfnamefont
  {J.}~\bibnamefont {Richter}}, \bibinfo {author} {\bibfnamefont
  {J.}~\bibnamefont {Malek}}, \bibinfo {author} {\bibfnamefont
  {A.}~\bibnamefont {Prokofiev}}, \bibinfo {author} {\bibfnamefont
  {W.}~\bibnamefont {Assmus}}, \bibinfo {author} {\bibfnamefont
  {S.}~\bibnamefont {Pujol}}, \bibinfo {author} {\bibfnamefont {J.-L.}\
  \bibnamefont {Raggazzoni}}, \bibinfo {author} {\bibfnamefont
  {H.}~\bibnamefont {Rakoto}}, \bibinfo {author} {\bibfnamefont
  {M.}~\bibnamefont {Rheinst{\"a}dter}}, \ and\ \bibinfo {author}
  {\bibfnamefont {H.~M.}\ \bibnamefont {R{\o}nnow}},\ }\href
  {http://stacks.iop.org/0295-5075/70/i=2/a=237} {\bibfield  {journal}
  {\bibinfo  {journal} {EPL (Europhysics Letters)}\ }\textbf {\bibinfo {volume}
  {70}},\ \bibinfo {pages} {237} (\bibinfo {year} {2005})}\BibitemShut
  {NoStop}%
\bibitem [{\citenamefont {Drechsler}\ \emph {et~al.}(2007)\citenamefont
  {Drechsler}, \citenamefont {Volkova}, \citenamefont {Vasiliev}, \citenamefont
  {Tristan}, \citenamefont {Richter}, \citenamefont {Schmitt}, \citenamefont
  {Rosner}, \citenamefont {M\'alek}, \citenamefont {Klingeler}, \citenamefont
  {Zvyagin},\ and\ \citenamefont {B\"uchner}}]{Drechsler:2007ve}%
  \BibitemOpen
  \bibfield  {author} {\bibinfo {author} {\bibfnamefont {S.-L.}\ \bibnamefont
  {Drechsler}}, \bibinfo {author} {\bibfnamefont {O.}~\bibnamefont {Volkova}},
  \bibinfo {author} {\bibfnamefont {A.~N.}\ \bibnamefont {Vasiliev}}, \bibinfo
  {author} {\bibfnamefont {N.}~\bibnamefont {Tristan}}, \bibinfo {author}
  {\bibfnamefont {J.}~\bibnamefont {Richter}}, \bibinfo {author} {\bibfnamefont
  {M.}~\bibnamefont {Schmitt}}, \bibinfo {author} {\bibfnamefont
  {H.}~\bibnamefont {Rosner}}, \bibinfo {author} {\bibfnamefont
  {J.}~\bibnamefont {M\'alek}}, \bibinfo {author} {\bibfnamefont
  {R.}~\bibnamefont {Klingeler}}, \bibinfo {author} {\bibfnamefont {A.~A.}\
  \bibnamefont {Zvyagin}}, \ and\ \bibinfo {author} {\bibfnamefont
  {B.}~\bibnamefont {B\"uchner}},\ }\href {\doibase
  10.1103/PhysRevLett.98.077202} {\bibfield  {journal} {\bibinfo  {journal}
  {Phys. Rev. Lett.}\ }\textbf {\bibinfo {volume} {98}},\ \bibinfo {pages}
  {077202} (\bibinfo {year} {2007})}\BibitemShut {NoStop}%
\bibitem [{\citenamefont {Park}\ \emph {et~al.}(2007)\citenamefont {Park},
  \citenamefont {Choi}, \citenamefont {Zhang},\ and\ \citenamefont
  {Cheong}}]{Park:2007yq}%
  \BibitemOpen
  \bibfield  {author} {\bibinfo {author} {\bibfnamefont {S.}~\bibnamefont
  {Park}}, \bibinfo {author} {\bibfnamefont {Y.~J.}\ \bibnamefont {Choi}},
  \bibinfo {author} {\bibfnamefont {C.~L.}\ \bibnamefont {Zhang}}, \ and\
  \bibinfo {author} {\bibfnamefont {S.-W.}\ \bibnamefont {Cheong}},\ }\href
  {\doibase 10.1103/PhysRevLett.98.057601} {\bibfield  {journal} {\bibinfo
  {journal} {Phys. Rev. Lett.}\ }\textbf {\bibinfo {volume} {98}},\ \bibinfo
  {pages} {057601} (\bibinfo {year} {2007})}\BibitemShut {NoStop}%
\bibitem [{\citenamefont {Cheong}\ and\ \citenamefont
  {Mostovoy}(2007)}]{Cheong:2007fk}%
  \BibitemOpen
  \bibfield  {author} {\bibinfo {author} {\bibfnamefont {S.-W.}\ \bibnamefont
  {Cheong}}\ and\ \bibinfo {author} {\bibfnamefont {M.}~\bibnamefont
  {Mostovoy}},\ }\href {http://dx.doi.org/10.1038/nmat1804} {\bibfield
  {journal} {\bibinfo  {journal} {Nat Mater}\ }\textbf {\bibinfo {volume}
  {6}},\ \bibinfo {pages} {13} (\bibinfo {year} {2007})}\BibitemShut {NoStop}%
\bibitem [{\citenamefont {White}(1992)}]{SWhite1992}%
  \BibitemOpen
  \bibfield  {author} {\bibinfo {author} {\bibfnamefont {S.~R.}\ \bibnamefont
  {White}},\ }\href {\doibase 10.1103/PhysRevLett.69.2863} {\bibfield
  {journal} {\bibinfo  {journal} {Phys. Rev. Lett.}\ }\textbf {\bibinfo
  {volume} {69}},\ \bibinfo {pages} {2863} (\bibinfo {year}
  {1992})}\BibitemShut {NoStop}%
\bibitem [{\citenamefont {White}(1993)}]{SWhite1993}%
  \BibitemOpen
  \bibfield  {author} {\bibinfo {author} {\bibfnamefont {S.~R.}\ \bibnamefont
  {White}},\ }\href {\doibase 10.1103/PhysRevB.48.10345} {\bibfield  {journal}
  {\bibinfo  {journal} {Phys. Rev. B}\ }\textbf {\bibinfo {volume} {48}},\
  \bibinfo {pages} {10345} (\bibinfo {year} {1993})}\BibitemShut {NoStop}%
\bibitem [{\citenamefont {Schollw\"ock}(2005)}]{Schollwock:2005qf}%
  \BibitemOpen
  \bibfield  {author} {\bibinfo {author} {\bibfnamefont {U.}~\bibnamefont
  {Schollw\"ock}},\ }\href {\doibase 10.1103/RevModPhys.77.259} {\bibfield
  {journal} {\bibinfo  {journal} {Rev. Mod. Phys.}\ }\textbf {\bibinfo {volume}
  {77}},\ \bibinfo {pages} {259} (\bibinfo {year} {2005})}\BibitemShut
  {NoStop}%
\bibitem [{\citenamefont {Hallberg}(2006)}]{Hallberg:2006bh}%
  \BibitemOpen
  \bibfield  {author} {\bibinfo {author} {\bibfnamefont {K.~A.}\ \bibnamefont
  {Hallberg}},\ }\href {\doibase 10.1080/00018730600766432} {\bibfield
  {journal} {\bibinfo  {journal} {Advances in Physics}\ }\textbf {\bibinfo
  {volume} {55}},\ \bibinfo {pages} {477} (\bibinfo {year} {2006})}\BibitemShut
  {NoStop}%
\bibitem [{\citenamefont {Bursill}\ \emph {et~al.}(1996)\citenamefont
  {Bursill}, \citenamefont {Xiang},\ and\ \citenamefont
  {Gehring}}]{Bursill:1996vn}%
  \BibitemOpen
  \bibfield  {author} {\bibinfo {author} {\bibfnamefont {R.~J.}\ \bibnamefont
  {Bursill}}, \bibinfo {author} {\bibfnamefont {T.}~\bibnamefont {Xiang}}, \
  and\ \bibinfo {author} {\bibfnamefont {G.~A.}\ \bibnamefont {Gehring}},\
  }\href {http://stacks.iop.org/0953-8984/8/i=40/a=003} {\bibfield  {journal}
  {\bibinfo  {journal} {Journal of Physics: Condensed Matter}\ }\textbf
  {\bibinfo {volume} {8}},\ \bibinfo {pages} {L583} (\bibinfo {year}
  {1996})}\BibitemShut {NoStop}%
\bibitem [{\citenamefont {Wang}\ and\ \citenamefont {Xiang}(1997)}]{XWang1997}%
  \BibitemOpen
  \bibfield  {author} {\bibinfo {author} {\bibfnamefont {X.}~\bibnamefont
  {Wang}}\ and\ \bibinfo {author} {\bibfnamefont {T.}~\bibnamefont {Xiang}},\
  }\href {\doibase 10.1103/PhysRevB.56.5061} {\bibfield  {journal} {\bibinfo
  {journal} {Phys. Rev. B}\ }\textbf {\bibinfo {volume} {56}},\ \bibinfo
  {pages} {5061} (\bibinfo {year} {1997})}\BibitemShut {NoStop}%
\bibitem [{\citenamefont {Xiang}(1998)}]{TXiang1998}%
  \BibitemOpen
  \bibfield  {author} {\bibinfo {author} {\bibfnamefont {T.}~\bibnamefont
  {Xiang}},\ }\href {\doibase 10.1103/PhysRevB.58.9142} {\bibfield  {journal}
  {\bibinfo  {journal} {Phys. Rev. B}\ }\textbf {\bibinfo {volume} {58}},\
  \bibinfo {pages} {9142} (\bibinfo {year} {1998})}\BibitemShut {NoStop}%
\bibitem [{\citenamefont {Naef}\ \emph {et~al.}(1999)\citenamefont {Naef},
  \citenamefont {Wang}, \citenamefont {Zotos},\ and\ \citenamefont {von~der
  Linden}}]{Naef:1999nx}%
  \BibitemOpen
  \bibfield  {author} {\bibinfo {author} {\bibfnamefont {F.}~\bibnamefont
  {Naef}}, \bibinfo {author} {\bibfnamefont {X.}~\bibnamefont {Wang}}, \bibinfo
  {author} {\bibfnamefont {X.}~\bibnamefont {Zotos}}, \ and\ \bibinfo {author}
  {\bibfnamefont {W.}~\bibnamefont {von~der Linden}},\ }\href {\doibase
  10.1103/PhysRevB.60.359} {\bibfield  {journal} {\bibinfo  {journal} {Phys.
  Rev. B}\ }\textbf {\bibinfo {volume} {60}},\ \bibinfo {pages} {359} (\bibinfo
  {year} {1999})}\BibitemShut {NoStop}%
\bibitem [{\citenamefont {Sirker}\ and\ \citenamefont
  {Kl{\"u}mper}(2002)}]{Sirker:2002kl}%
  \BibitemOpen
  \bibfield  {author} {\bibinfo {author} {\bibfnamefont {J.}~\bibnamefont
  {Sirker}}\ and\ \bibinfo {author} {\bibfnamefont {A.}~\bibnamefont
  {Kl{\"u}mper}},\ }\href {http://stacks.iop.org/0295-5075/60/i=2/a=262}
  {\bibfield  {journal} {\bibinfo  {journal} {EPL (Europhysics Letters)}\
  }\textbf {\bibinfo {volume} {60}},\ \bibinfo {pages} {262} (\bibinfo {year}
  {2002})}\BibitemShut {NoStop}%
\bibitem [{\citenamefont {Sirker}(2004)}]{Sirker:2004oq}%
  \BibitemOpen
  \bibfield  {author} {\bibinfo {author} {\bibfnamefont {J.}~\bibnamefont
  {Sirker}},\ }\href {\doibase 10.1103/PhysRevB.69.104428} {\bibfield
  {journal} {\bibinfo  {journal} {Phys. Rev. B}\ }\textbf {\bibinfo {volume}
  {69}},\ \bibinfo {pages} {104428} (\bibinfo {year} {2004})}\BibitemShut
  {NoStop}%
\bibitem [{\citenamefont {Lu}\ \emph {et~al.}(2006)\citenamefont {Lu},
  \citenamefont {Wang}, \citenamefont {Qin},\ and\ \citenamefont
  {Xiang}}]{Lu:2006fk}%
  \BibitemOpen
  \bibfield  {author} {\bibinfo {author} {\bibfnamefont {H.~T.}\ \bibnamefont
  {Lu}}, \bibinfo {author} {\bibfnamefont {Y.~J.}\ \bibnamefont {Wang}},
  \bibinfo {author} {\bibfnamefont {S.}~\bibnamefont {Qin}}, \ and\ \bibinfo
  {author} {\bibfnamefont {T.}~\bibnamefont {Xiang}},\ }\href {\doibase
  10.1103/PhysRevB.74.134425} {\bibfield  {journal} {\bibinfo  {journal} {Phys.
  Rev. B}\ }\textbf {\bibinfo {volume} {74}},\ \bibinfo {pages} {134425}
  (\bibinfo {year} {2006})}\BibitemShut {NoStop}%
\bibitem [{\citenamefont {Sota}\ and\ \citenamefont
  {Tohyama}(2010)}]{Sota:2010fk}%
  \BibitemOpen
  \bibfield  {author} {\bibinfo {author} {\bibfnamefont {S.}~\bibnamefont
  {Sota}}\ and\ \bibinfo {author} {\bibfnamefont {T.}~\bibnamefont {Tohyama}},\
  }\href {http://stacks.iop.org/1742-6596/200/i=1/a=012191} {\bibfield
  {journal} {\bibinfo  {journal} {Journal of Physics: Conference Series}\
  }\textbf {\bibinfo {volume} {200}},\ \bibinfo {pages} {012191} (\bibinfo
  {year} {2010})}\BibitemShut {NoStop}%
\bibitem [{\citenamefont {Huang}(2011{\natexlab{a}})}]{YHuang2011a}%
  \BibitemOpen
  \bibfield  {author} {\bibinfo {author} {\bibfnamefont {Y.-K.}\ \bibnamefont
  {Huang}},\ }\href {\doibase 10.1103/PhysRevE.83.036702} {\bibfield  {journal}
  {\bibinfo  {journal} {Phys. Rev. E}\ }\textbf {\bibinfo {volume} {83}},\
  \bibinfo {pages} {036702} (\bibinfo {year} {2011}{\natexlab{a}})}\BibitemShut
  {NoStop}%
\bibitem [{\citenamefont {Huang}(2011{\natexlab{b}})}]{YHuang2011b}%
  \BibitemOpen
  \bibfield  {author} {\bibinfo {author} {\bibfnamefont {Y.-K.}\ \bibnamefont
  {Huang}},\ }\href {http://stacks.iop.org/1742-5468/2011/i=07/a=P07003}
  {\bibfield  {journal} {\bibinfo  {journal} {J. Stat. Mech.: Theory Exp.}\
  }\textbf {\bibinfo {volume} {2011}},\ \bibinfo {pages} {P07003} (\bibinfo
  {year} {2011}{\natexlab{b}})}\BibitemShut {NoStop}%
\bibitem [{\citenamefont {Trotter}(1958)}]{HTrotter1958}%
  \BibitemOpen
  \bibfield  {author} {\bibinfo {author} {\bibfnamefont {H.~F.}\ \bibnamefont
  {Trotter}},\ }\href@noop {} {\bibfield  {journal} {\bibinfo  {journal} {Proc.
  Amer. Math. Soc.}\ }\textbf {\bibinfo {volume} {10}},\ \bibinfo {pages} {545}
  (\bibinfo {year} {1958})}\BibitemShut {NoStop}%
\bibitem [{\citenamefont {Suzuki}(1976)}]{MSuzuki1976}%
  \BibitemOpen
  \bibfield  {author} {\bibinfo {author} {\bibfnamefont {M.}~\bibnamefont
  {Suzuki}},\ }\href {http://dx.doi.org/10.1007/BF01609348} {\bibfield
  {journal} {\bibinfo  {journal} {Common. Math. Phys.}\ }\textbf {\bibinfo
  {volume} {51}},\ \bibinfo {pages} {183} (\bibinfo {year} {1976})},\ \bibinfo
  {note} {10.1007/BF01609348}\BibitemShut {NoStop}%
\bibitem [{\citenamefont {Suzuki}(1985)}]{MSuzuki1985}%
  \BibitemOpen
  \bibfield  {author} {\bibinfo {author} {\bibfnamefont {M.}~\bibnamefont
  {Suzuki}},\ }\href {\doibase 10.1103/PhysRevB.31.2957} {\bibfield  {journal}
  {\bibinfo  {journal} {Phys. Rev. B}\ }\textbf {\bibinfo {volume} {31}},\
  \bibinfo {pages} {2957} (\bibinfo {year} {1985})}\BibitemShut {NoStop}%
\bibitem [{\citenamefont {Nomura}\ and\ \citenamefont
  {Yamada}(1991)}]{Nomura:1991uq}%
  \BibitemOpen
  \bibfield  {author} {\bibinfo {author} {\bibfnamefont {K.}~\bibnamefont
  {Nomura}}\ and\ \bibinfo {author} {\bibfnamefont {M.}~\bibnamefont
  {Yamada}},\ }\href {\doibase 10.1103/PhysRevB.43.8217} {\bibfield  {journal}
  {\bibinfo  {journal} {Phys. Rev. B}\ }\textbf {\bibinfo {volume} {43}},\
  \bibinfo {pages} {8217} (\bibinfo {year} {1991})}\BibitemShut {NoStop}%
\bibitem [{\citenamefont {Kl{\"u}mper}(1993)}]{Klumper:1993ly}%
  \BibitemOpen
  \bibfield  {author} {\bibinfo {author} {\bibfnamefont {A.}~\bibnamefont
  {Kl{\"u}mper}},\ }\href {http://dx.doi.org/10.1007/BF01316831} {\bibfield
  {journal} {\bibinfo  {journal} {Zeitschrift f{\"u}r Physik B Condensed
  Matter}\ }\textbf {\bibinfo {volume} {91}},\ \bibinfo {pages} {507} (\bibinfo
  {year} {1993})},\ \bibinfo {note} {10.1007/BF01316831}\BibitemShut {NoStop}%
\bibitem [{\citenamefont {Kl{\"u}mper}(2004)}]{Klumper:2004zr}%
  \BibitemOpen
  \bibfield  {author} {\bibinfo {author} {\bibfnamefont {A.}~\bibnamefont
  {Kl{\"u}mper}},\ }in\ \href {http://dx.doi.org/} {\emph {\bibinfo {booktitle}
  {Quantum Magnetism}}},\ \bibinfo {series} {Lecture Notes in Physics}, Vol.\
  \bibinfo {volume} {645}\ (\bibinfo  {publisher} {Springer Berlin /
  Heidelberg},\ \bibinfo {year} {2004})\ pp.\ \bibinfo {pages}
  {349--379}\BibitemShut {NoStop}%
\bibitem [{\citenamefont {Sirker}(2002)}]{Sirker:2002ys}%
  \BibitemOpen
  \bibfield  {author} {\bibinfo {author} {\bibfnamefont {J.}~\bibnamefont
  {Sirker}},\ }\emph {\bibinfo {title} {Transfer matrix approach to
  thermodynamics and dynamics of one-dimensional quantum systems}},\ \href@noop
  {} {Ph.D. thesis},\ \bibinfo  {school} {Universit\"{a}t Dortmund} (\bibinfo
  {year} {2002})\BibitemShut {NoStop}%
\bibitem [{\citenamefont {Sirker}\ and\ \citenamefont
  {Kl\"umper}(2005)}]{Sirker:2005tg}%
  \BibitemOpen
  \bibfield  {author} {\bibinfo {author} {\bibfnamefont {J.}~\bibnamefont
  {Sirker}}\ and\ \bibinfo {author} {\bibfnamefont {A.}~\bibnamefont
  {Kl\"umper}},\ }\href {\doibase 10.1103/PhysRevB.71.241101} {\bibfield
  {journal} {\bibinfo  {journal} {Phys. Rev. B}\ }\textbf {\bibinfo {volume}
  {71}},\ \bibinfo {pages} {241101} (\bibinfo {year} {2005})}\BibitemShut
  {NoStop}%
\end{thebibliography}%

\end{document}